\documentclass[a4paper, amsfonts, amssymb, amsmath, reprint, showkeys, nofootinbib, twoside]{revtex4-2}
\usepackage[english]{babel}
\usepackage[utf8]{inputenc}
\usepackage{amsmath}
\usepackage{dsfont}
\usepackage{csquotes}
\usepackage{graphicx}
\usepackage[justification=raggedright, singlelinecheck=false,
            labelsep=period]{caption}

\AtBeginDocument{}
\usepackage{subcaption}
\usepackage{braket}
\usepackage{enumitem}
\usepackage{tikz}
\usepackage{tkz-berge}
\usepackage{xcolor}
\usepackage[mode=buildnew]{standalone}
\usepackage{float}
\usepackage{placeins}
\usepackage[colorinlistoftodos, color=green!40, prependcaption]{todonotes}
\usepackage[pdftex, pdftitle={Article}, pdfauthor={Author}]{hyperref} 

\usepackage[colorinlistoftodos, color=green!40, prependcaption]{todonotes}
\usepackage{todonotes}

\begin{document}
\title{James-Stein estimation for quantum sensing schemes}

\author{Luke Alexander Rhodes$^{1}$, Sean William Moore$^{2}$ and Jacob A. Dunningham$^{1}$}
    \affiliation{$^{1}$Department of Physics and Astronomy, University of Sussex, Brighton BN1 9QH, United Kingdom \\
    $^{2}$Sorbonne Universit\'e, CNRS, LIP6, F-75005 Paris, France}


\begin{abstract}
Quantum metrology protocols typically consist of four steps: state preparation, evolution, measurement, and data processing. Often, the first three steps are prioritised when designing a scheme as they contain all the quantum elements. The data analysis is generally considered an add-on with an implicit assumption that this step is well behaved and so standard data techniques can be applied. However, the situation can be more nuanced, such as when the available data are limited. In limited-data quantum metrology the choice of data analysis technique and cost function of the estimator is of great importance, and a reliable prior distribution of the unknown parameters is required for Bayesian analysis. An interesting question is what we should do when no such prior is available. In this work, we consider how the James-Stein estimator 
can give significant advantages when measuring multiple unknown parameters with limited data and, importantly, does not require any  prior distribution. We demonstrate the advantage by applying this methodology to simple quantum metrology schemes.
\end{abstract}


\maketitle

\section{Introduction}
Quantum metrology and sensing exploit quantum mechanical resources, such as entanglement and squeezing, to enhance measurement precision and detection sensitivity beyond the limits achievable with classical systems~\cite{Giovannetti2006}.  These techniques represent one of the most mature areas of quantum technology and have applications ranging from gravitational-wave detection~\cite{LIGO2013}, atomic clocks and frequency standards~\cite{Ludlow2015}, and magnetic and electric field sensing~\cite{Degen2017}, to imaging~\cite{Moreau2019}, navigation~\cite{Kasevich2002}, and searches for physics beyond the Standard Model~\cite{Graham2013}. Quantum metrology protocols typically involve four stages: preparation of a suitable probe state, evolution of the probe to encode the parameters of interest, measurement of the encoded quantum state, and classical processing of the resulting measurement data~\cite{Pezze2017}. Quantum metrology generally focuses on the first three stages using powerful tools such as Fisher information~\cite{Braunstein1994}.
These stages are carefully optimised together since they are interconnected, e.g. the choice of quantum state is intimately linked to the measurement that will be made. 
The last stage is often overlooked since it does not contain the quantum part of the scheme and is solely concerned with classical data. We argue that it should be taken seriously in the overall design of quantum metrology schemes and that the data processing strategy can have a big impact on overall performance. Here we consider this in the particular context of using the James-Stein estimator \cite{Stein61} to improve the estimation of multiple parameters.

A particularly active area of recent research in multi-parameter quantum metrology is that of distributed quantum sensing networks, in which spatially separated quantum sensors are correlated or entangled to estimate global properties of a field or environment~\cite{Proctor2018,Proctor2017,Eldredge2018}. Such architectures are especially advantageous when the goal is to estimate a function of parameters distributed between different nodes, rather than the individual parameters themselves~\cite{Proctor2018}. Distributed sensing has potential applications in precision navigation, gravitational and magnetic field sensing, astronomy, and geodesy~\cite{Eldredge2018,Zhuang2020,Ge2021}. It has also been proposed as a route towards a secure and highly precise global quantum clock network~\cite{Komar2014}, with significant experimental progress towards networked optical clock systems already being demonstrated~\cite{Nichol2022}.

As quantum metrology matures, emphasis has been placed on extending it to new regimes of applicability, such as when data are limited~\cite{Rubio2020, Rubio2019}, or combining it with other quantum technologies such as quantum communications in secure sensing schemes~\cite{Moore2023,Hassani2025}. Another direction has been to make schemes more practical to implement, e.g. by mitigating the effects of noise through error correction~\cite{Kessler2014, Zhou2018}, improving noise diagnostics, or using more robust quantum states~\cite{Dorner2009}. Experimental accessibility can also be improved by investigating what can be achieved without the use of entangled states~\cite{Braun2018, Moore2023}. This helps with the practicality of the state preparation stage, especially for networks with many nodes, and also improves the state fidelity. Such an approach has the added advantage that separable states have been shown to be exponentially more efficient in checking for noise or eavesdroppers in certain secure distributed sensing protocols~\cite{Moore2025}


The focus of literature for multi-parameter quantum metrology has been on the estimation of more than one parameter with a single probe system, estimation with unknown noise parameters and the estimation of functions of parameters, usually a single function. In this paper, we consider a different perspective: how can we improve the average quality of an estimation of several independent parameters accessed by independent systems. In particular, we will focus on the James-Stein estimator, which has only recently begun to be applied to quantum metrology for Gaussian states~\cite{Salmon2024}. Here we show that significant advantages in precision can be achieved over the standard maximum likelihood estimator when estimating three or more normally distributed parameters. This can be achieved in a straightforward manner without increasing the quantum resource cost. 

We demonstrate how these methods can be applied to discrete variable quantum metrology problems as well as  systems with quantum-enhanced measurements. Such a method could also be extended to a broad range of other schemes since it simply relies on changing post-processing techniques rather than relying on a particular setup or requiring additional experimental resources.

The structure of this paper is as follows. Section~\ref{sec:JSE}  introduces the James-Stein estimator (JSE) and shows how it improves upon the maximum likelihood estimator (MLE) in terms of the average mean square error (MSE) on three or more independent normally-distributed parameters. Section~\ref{sec:JSEext} sets out extensions to the scheme to improve its performance and deal with the practical limitation of not knowing the mean of the parameters a priori. Sections~\ref{sec:quantumSensing} and~\ref{sec:SensingStats} apply these ideas to a standard quantum sensing scheme with separable qubits and consider the requirements for effective application of  shrinkage estimation techniques. Section~\ref{sec:Simulations}  presents simulations of the estimation of several phases using the JSE and, finally, Section~\ref{sec:quantumEnhanced} shows how these methods can be combined with quantum metrological techniques using  entangled states to gain a further advantage in measurement precision.

\section{The James-Stein estimator} \label{sec:JSE}
\begin{figure}[h]
    \centering
    \includegraphics[width=0.45\textwidth]{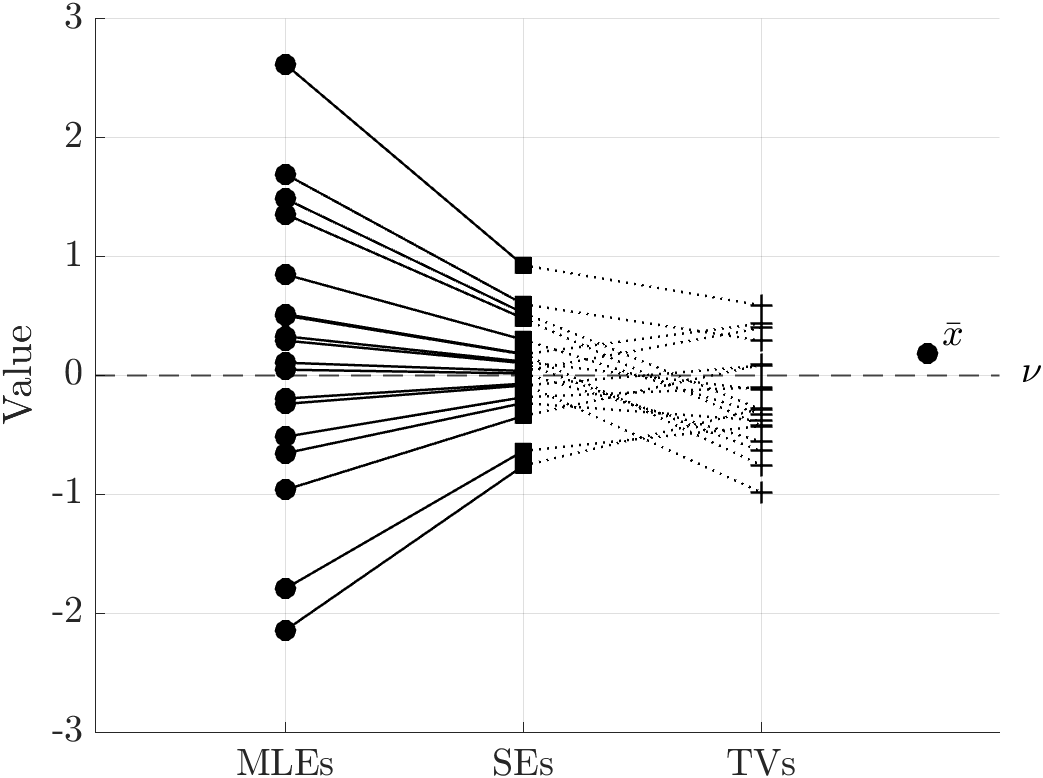}
    \caption{The idea behind shrinkage estimators (SEs). The initial MLEs are scaled towards some central point $\nu$ (or if this is not known, towards the mean of the MLEs, $\bar{x}$) in an attempt to \enquote*{shrink} some, or all, of the components closer to the underlying true values. Note that it is possible for some of the individual components to be moved further away.}
    \label{fig:Figure1}
\end{figure}

To introduce the James-Stein estimator in a simple setting, we follow~\cite{Efron75}. Suppose that we observe
\[
x_i\mid \phi_i \overset{\text{ind}}{\sim} \mathcal{N}(\phi_i, \sigma^2) \quad \text{where}\ i = 1,2,...,d \quad \text{and}\ d \geq 3.
\]
where each $x_i$ is the maximum likelihood estimate for the corresponding $\phi_i$. Note that here we are assuming that the variance of each likelihood function is equal and known. However, the model is surprisingly robust to violations in these assumptions and in practice it is safe to estimate these variances using sample data with negligible loss in performance. The unknown vector of means 
\[
\boldsymbol{\phi} = (\phi_1, \phi_2, ..., \phi_d) 
\]
is most commonly estimated in networked quantum metrology by using the vector of maximum likelihood estimators (MLEs), given by 
\[
\boldsymbol{\delta}^\text{MLE}(\boldsymbol{x}) = \boldsymbol{x} = (x_1,x_2,...,x_d).
\]
If the loss function is chosen to be the sum of squared errors in each independent component, that is, 
\[
L(\boldsymbol{\phi},\boldsymbol{\delta}) = \|\boldsymbol{\phi}-\boldsymbol{\delta}\|^2
\]
and risk is defined to be the expectation value (or average over many repeat experiments) of loss for a fixed $\boldsymbol{\phi}$, then 
\[
R(\boldsymbol{\phi},\boldsymbol{\delta}) = \mathbb{E}_{\boldsymbol{\phi}}\|\boldsymbol{\phi}-\boldsymbol{\delta}\|^2.
\]
This is the same as the mean-squared error (MSE) summed across the independent components, and so we will use the terms risk and MSE interchangeably. Through linearity of expectation and summation, the risk of the MLE is given by
\begin{equation}
R(\boldsymbol{\phi},\boldsymbol{\delta}^\text{MLE}) = d\sigma^2,
\label{eq:MLErisk}
\end{equation}
which is independent of $\boldsymbol{\phi}$ and scales linearly with the number of parameters, $d$. Since 
\begin{equation}
\mathbb{E}_{\boldsymbol{\phi}}\!\left[\|\boldsymbol{x}\|^2\right]
= \|\boldsymbol{\phi}\|^2 + d\sigma^2,
\label{eq:ExpectationSumX^2}
\end{equation}
we see that the additional term $d\sigma^2$ means that, as an estimator of $\boldsymbol{\phi}$, the magnitude of the vector $\boldsymbol{x}$ is too large. If we visualise each $x_i$ and $\phi_i$ as points on a number line as in~\cite{Casella85}, Eq.~(2) indicates that these MLEs will be over-dispersed relative to the unknown true values (TVs) and that we may be able to intentionally bias the MLEs towards some central origin, $\nu$, so as to produce a new set of estimates that are closer on average to the true values. This idea is illustrated in Fig.~\ref{fig:Figure1}. Alternatively, if the central origin is unknown, we could compute the group mean of the MLEs themselves, $\bar{x}$, and shrink our values towards that. We will consider this case in Sec.~\ref{sec:JSEext}. 

We now turn our attention to the form of estimators that can correct for the over-dispersion and \textit{dominate} the MLE. An estimator, $\delta_1$, is said to dominate another, $\delta_2$, if 
\[
R(\phi,\delta_1) \leq R(\phi,\delta_2)
\]
for all $\phi$~\cite{LehmannV1}. It was first demonstrated in~\cite{Stein56} that estimators exist which dominate the MLE in the multivariate normal setting, and it was shown in~\cite{Stein61} that 
\begin{equation}
\boldsymbol{\delta}^\text{JSE}(\boldsymbol{x}) = \left(1-\frac{(d-2)\sigma^2}{\|\boldsymbol{x}\|^2}\right)\boldsymbol{x},
\label{eq:JS0}
\end{equation}
which scales the vector of MLEs towards zero, dominates the MLE. This is now known as the James-Stein estimator and this particular form is best suited to the case where all $\phi_i$ are believed to take values near zero. It can be readily generalised to the case for an arbitrary choice of $\nu$ by using 
\begin{equation}
\boldsymbol{\delta}^\text{JSE}(\boldsymbol{x},\boldsymbol{\nu}) = \boldsymbol{\nu}+\left(1-\frac{(d-2)\sigma^2}{\|\boldsymbol{x}-\boldsymbol{\nu}\|^2}\right)(\boldsymbol{x}-\boldsymbol{\nu})
\label{eq:JSnu}
\end{equation}
which acts as a linear translation of the problem. For simplicity, in the rest of this paper we will assume that  
\[
\nu_1 = \nu_2 = ... =\nu_d = \nu,
\]
suggestive of our belief that all the $\phi_i$ are clustered around the same scalar point in the parameter space, but it is straightforward to scale different components towards different points if needed. 

\begin{figure}[t]
    \centering
    \includegraphics[width=\columnwidth]{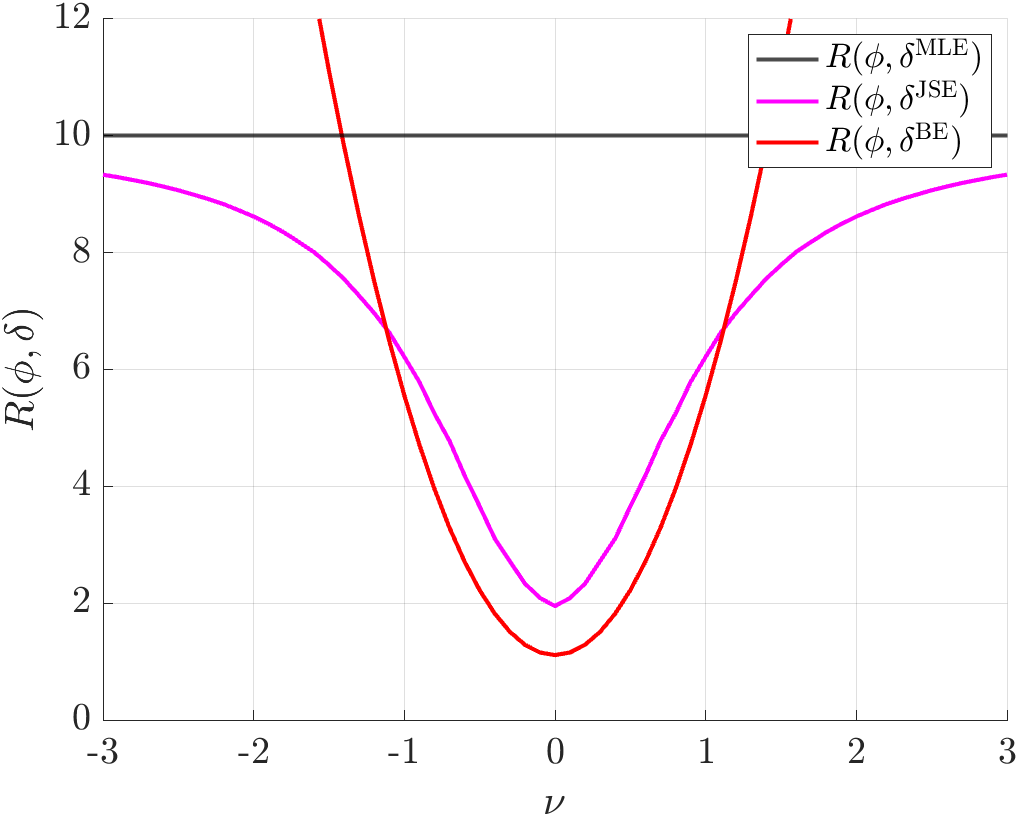}
    \caption{MSE as a function of $\nu$ for the MLE, JSE and Bayes estimator with $\|\boldsymbol{\phi}\|^2=0$,  $d=10$, $\sigma = 1$ and $\tau = 1/\sqrt{2}$ on an interval of the real line.}
    \label{fig:Figure2}
\end{figure}

The risk function of $\boldsymbol{\delta}^\text{JSE}(\boldsymbol{x},\boldsymbol{\nu})$ is given in~\cite{FSW} as
\begin{equation}
R(\boldsymbol{\phi},\boldsymbol{\delta}^\text{JSE}) = d\sigma^2 - (d-2)^2\sigma^4\mathbb{E}_{\boldsymbol{\phi}}\left[\frac{1}{\|\boldsymbol{x}-\boldsymbol{\nu}\|^2}\right].
\label{eq:JSnurisk}
\end{equation}
and it can be shown that the risk obtained by using the James-Stein estimator gets smaller as $d \to \infty$. In other words, the improvement is most pronounced for higher dimensional problems. The second term in \eqref{eq:JSnurisk} is  proportional to $\sigma^4$ and so 
\begin{equation}
R(\boldsymbol{\phi},\boldsymbol{\delta}^\text{JSE}) \to R(\boldsymbol{\phi},\boldsymbol{\delta}^\text{MLE}) \quad \text{as } \sigma^2\to 0. 
\label{JSrisktendstoMLErisk}
\end{equation}
implying that the JSE is particularly useful when the variance is high. From~\cite{FSW}, the lowest attainable value for \eqref{eq:JSnurisk} occurs when our choice of the vector $\boldsymbol{\nu}$ is exactly the vector of true values $\boldsymbol{\phi}$, in which case we observe 
\begin{equation}
 R(\boldsymbol{\phi},\boldsymbol{\delta}^\text{JSE}) = 2\sigma^2,
    \label{JSRiskAtZero}
\end{equation}
which constitutes a significant improvement over the MLE, especially for large $d$ (compare with \eqref{eq:MLErisk}). Interestingly, note that we could in fact specify \textit{any} choice of shrinkage target $\boldsymbol{\nu}$ and we would never make the risk worse than that of the MLE. This is because
\[
\mathbb{E}_{\boldsymbol{\phi}}\left[\frac{1}{\|\boldsymbol{x}-\boldsymbol{\nu}\|^2}\right] \to 0 \quad \text{as } \|\boldsymbol{x}-\boldsymbol{\nu}\|^2 \to \infty,
\]
and so
\begin{equation}  R(\boldsymbol{\phi},\boldsymbol{\delta}^\text{JSE}) \to R(\boldsymbol{\phi},\boldsymbol{\delta}^\text{MLE})\quad \text{as } \|\boldsymbol{x}-\boldsymbol{\nu}\|^2 \to \infty,
    \label{JSRiskAtInfinity}
\end{equation}
hence satisfying the dominance condition.
This behaviour is observed in Fig.~\ref{fig:Figure2}, where \eqref{eq:JSnurisk} is plotted as a function $\nu$ for the case $\phi_i = 0$. Importantly, the risk of the JSE is never higher than that of the MLE, irrespective of the choice of $\nu$ or the value of $\phi_i$, and can be much lower. 

This is in stark contrast to a fully Bayesian approach. For this, a prior distribution of the TVs needs to be assumed, and here we take this to be,
\begin{equation}
    \phi_i \sim  \mathcal{N}(\nu, \tau^2),
    \label{TVdistributionNormal}
\end{equation}
where, as before, $i = 1,2,...d$, $d\geq 3$ and all components of the vector $\boldsymbol{\nu}$ are identical. The Bayes estimator (BE) for this multivariate normal likelihood-normal prior model is the posterior mean~\cite{Carlin}, given by 
\[
\boldsymbol{\delta}^\text{BE}(\boldsymbol{x},\boldsymbol{\nu}) = \boldsymbol{\nu}+\left(1-B\right)(\boldsymbol{x}-\boldsymbol{\nu}), 
\]
where
\[
B = \frac{\sigma^2}{\tau^2+\sigma^2}
\]
and has risk function 
\begin{equation} R(\boldsymbol{\phi},\boldsymbol{\delta}^\text{BE}) = (1-B)^2d\sigma^2 +B^2\|\boldsymbol{\phi}-\boldsymbol{\nu}\|^2. 
   \label{eq:BErisk}
\end{equation}

The BE is compared with the JSE in Figure~\ref{fig:Figure2}. We see that reliable prior information (i.e  when the prior is centred close to the TVs) can greatly reduce the risk in the estimate, even more so than the JSE, but can greatly increase the risk if the prior is poor. Given that the values of $\phi_i$ are unknown, there is a clear drawback to using the Bayesian method, since it requires the $\phi_i$ to be known to a high degree of confidence in the first place. Although fully frequentist in nature, the JSE has often been framed as an \enquote*{empirical Bayes} estimator \cite{Efron75, Casella85}, since it serves as a form of compromise between the fully Bayesian method and the fully frequentist MLE. The JSE mimics the behaviour of the BE when it is beneficial to do so, in the central region of the parameter space, but tends to the MLE in the extremities. This acts as insurance against the large blow-up in MSE that is possible when using the Bayesian method. 

It is worth stating that these results are based on expectation values and so hold on average. For a single run of an estimation procedure or for any of the individual components $\phi_i$, the MLE or BE could provide an estimate with lower squared-loss than the JSE. The JSE is thus of most use when long-run performance averaged across all the $\phi_i$ is of principal interest, as opposed to that for an individual component.  

\section{Extensions to the James-Stein Estimator} \label{sec:JSEext}
We now discuss the two most common adaptations to the JSE. Thus far we have simply specified an arbitrary vector $\boldsymbol{\nu}$ and have shown that even if we get it wrong we still do at least as well as the MLE. What if instead we chose to let the data decide the point to shrink towards? This correction, due to Lindley~\cite{Lindley62}, shrinks towards the group mean of the MLEs, and results in what is known as the Lindley estimator, which we will denote JSL,
\begin{equation}
 \boldsymbol{\delta}^\text{JSL}(\boldsymbol{x}) = \boldsymbol{\bar{x}}+\left(1-\frac{(d-3)\sigma^2}{\|\boldsymbol{x}-\boldsymbol{\bar{x}}\|^2}\right)(\boldsymbol{x}-\boldsymbol{\bar{x}}),
\label{eq:JSL}   
\end{equation}
where 
\[
\boldsymbol{\bar{x}} = \left(\frac{1}{d}\sum_{i=1}^{d}{x_i}\right)\mathds{1},
\]
i.e. a vector, the same length as $\boldsymbol{x}$, where each component is equal to the sample mean of the MLEs. The corresponding risk function~\cite{FSW} is given by 
\begin{equation}
  R(\boldsymbol{\phi},\boldsymbol{\delta}^\text{JSL}) = d\sigma^2 - (d-3)^2\sigma^4\mathbb{E}_{\boldsymbol{\phi}}\left[\frac{1}{\|\boldsymbol{x}-\boldsymbol{\bar{x}}\|^2}\right]
\label{eq:JSLrisk}  
\end{equation}
A key difference from the JSE is that we now require $d \geq 4$ in order to get a reduction in risk; this arises since we have estimated an additional variable (the point to shrink towards), so a degree of freedom is lost in the derivation of the estimator. For the Lindley estimator, in the case where all the $\phi_i$ are identical,
\[
R(\boldsymbol{\phi},\boldsymbol{\delta}^\text{JSL}) = 3\sigma^2
\]
which still offers a large potential improvement over the MLE~\cite{FSW}, but no so great as ~\eqref{eq:JSnu} (see Eq.~\eqref{JSRiskAtZero}). 

Another useful adaptation to the JSE is the so-called positive-part condition, which we denote JSE+. From \eqref{eq:JSnu}, we can see that if 
\[
(d-2)\sigma^2 > \|\boldsymbol{x}-\boldsymbol{\nu}\|^2
\]
then the estimator shrinks $\boldsymbol{x}$ the wrong way, i.e. away from the point we believe the true values are clustered around. The positive-part condition~\cite{LehmannV2} protects against this by insisting that \eqref{eq:JSnu} be re-written as 
\begin{equation}
    \boldsymbol{\delta}^\text{JSE+}(\boldsymbol{x},\boldsymbol{\nu}) = \boldsymbol{\nu}+\left(1-m_{\boldsymbol{\nu}}\right)(\boldsymbol{x}-\boldsymbol\nu),
\label{eq:JSnuPP}
\end{equation}
where 
\begin{equation}
    m_{\boldsymbol{\nu}} =
    \begin{cases}
    \frac{(d-2)\sigma^2}{\|\boldsymbol{x}-\boldsymbol{\nu}\|^2} & \text{if } (d-2)\sigma^2 < \|\boldsymbol{x}-\boldsymbol{\nu}\|^2 \\
    1 & \text{if } (d-2)\sigma^2 \geq \|\boldsymbol{x}-\boldsymbol{\nu}\|^2.
    \end{cases}
\end{equation}

\begin{figure}[h]
    \centering
    \hspace*{-40pt}
    \includegraphics[width=\columnwidth,height=0.33\textheight,keepaspectratio]{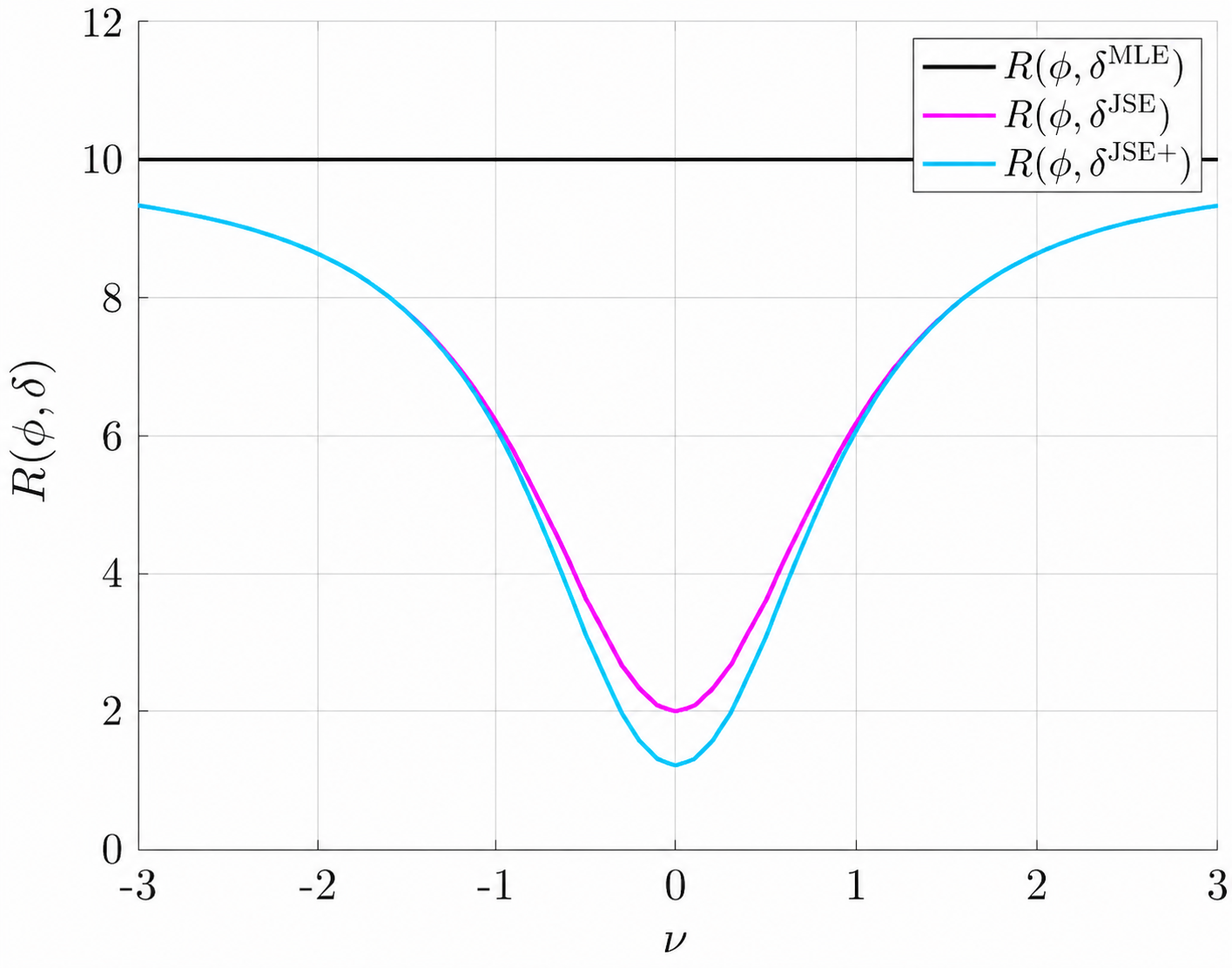}
    \medskip
    \hspace*{-40pt}
    \includegraphics[width=\columnwidth,height=0.33\textheight,keepaspectratio]{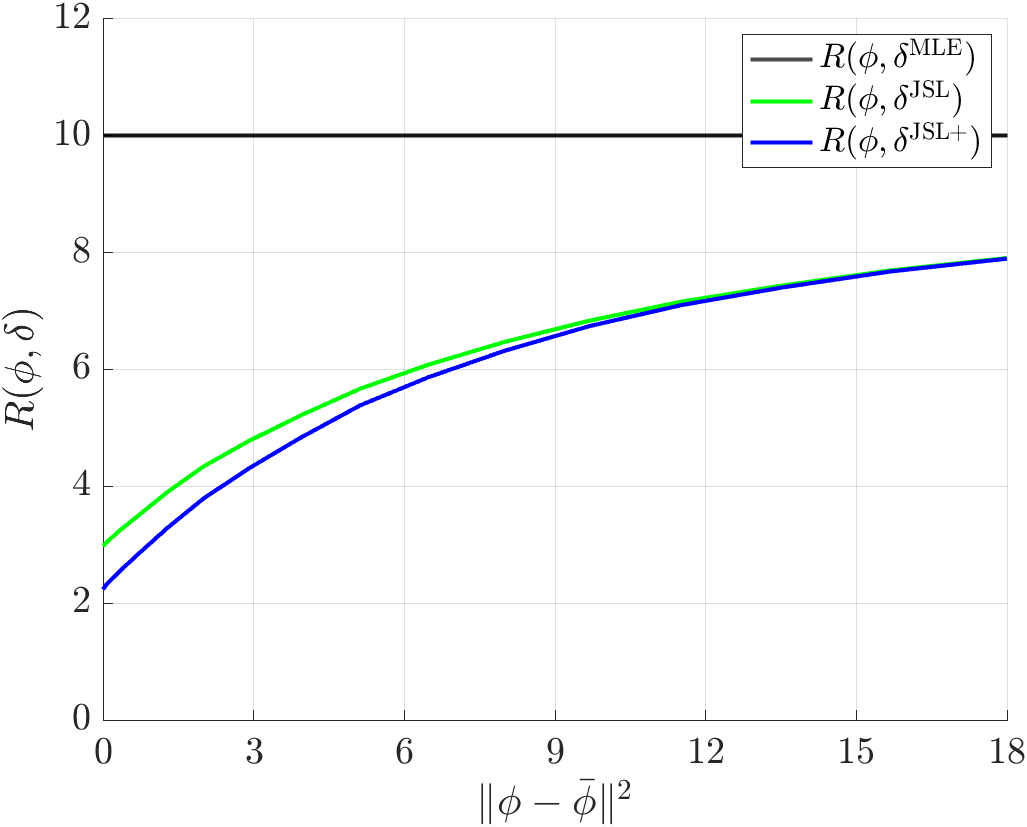}
    \caption{
    The positive-part versions dominate the standard estimators. The upper plot compares the risk of JSE+, given by \eqref{eq:JSnuPPrisk} to the risk of the standard JSE given by \eqref{eq:JSnurisk}. The lower plot compares the risk of JSL and JSL+, given by  \eqref{eq:JSLrisk}  and \eqref{eq:JSLPPrisk} respectively. For both parts we have taken $\sigma = 1$.}
    \label{fig:PPriskplots}
\end{figure}

The risk function for $\boldsymbol{\delta}^\text{JSE+}(\boldsymbol{x},\boldsymbol{\nu})$ is given by 
\begin{equation}
    \begin{aligned}
    R\bigl(\boldsymbol{\phi},
    \boldsymbol{\delta}^{\text{JSE+}}\bigr)
    &= \mathbb{E}_{\boldsymbol{\phi}}\Bigl[
    d\sigma^2
    + m_{\boldsymbol{\nu}}
      \lVert \boldsymbol{x}-\boldsymbol{\nu} \rVert^2 \\
    &\quad - 2\sigma^2 (d-2)\,
    m_{\boldsymbol{\nu}}
    - 4\sigma^2
    \mathbb{I}\!\left[m_{\boldsymbol{\nu}}=1\right]
    \Bigr],
    \end{aligned}
    \label{eq:JSnuPPrisk}
\end{equation}
where $\mathbb{I} = [\,\cdot\,]$ is an indicator function that equals one if the argument within the brackets is satisfied, and zero otherwise. 
An analogous correction can also be made to the Lindley version of the estimator~\cite{LehmannV2} (which we denote JSL+) so that \eqref{eq:JSL} becomes
\begin{equation}
    \boldsymbol{\delta}^\text{JSL+}(\boldsymbol{x}) = \boldsymbol{\bar{x}}+\left(1-m_{\boldsymbol{\bar{x}}}\right)(\boldsymbol{x}-\boldsymbol{\bar{x}})
\label{eq:JSLPP}
\end{equation}
where 
\begin{equation}
    m_{\boldsymbol{\bar{x}}} =
    \begin{cases}
    \frac{(d-3)\sigma^2}{\|\boldsymbol{x}-\boldsymbol{\bar{x}}\|^2} & \text{if } (d-3)\sigma^2 < \|\boldsymbol{x}-\boldsymbol{\bar{x}}\|^2 \\
    1 & \text{if } (d-3)\sigma^2 \geq \|\boldsymbol{x}-\boldsymbol{\bar{x}}\|^2
    \end{cases}
\end{equation}
and \eqref{eq:JSLrisk} becomes 
\begin{equation}
    \begin{aligned}
    R\bigl(\boldsymbol{\phi},
    \boldsymbol{\delta}^{\text{JSL+}}\bigr)
    &= \mathbb{E}_{\boldsymbol{\phi}}\Bigl[
    d\sigma^2
    + m_{\boldsymbol{\bar{x}}}
      \lVert \boldsymbol{x}-\boldsymbol{\bar{x}} \rVert^2 \\
    &\quad - 2\sigma^2 (d-3)\,
    m_{\boldsymbol{\bar{x}}}
    - 4\sigma^2
    \mathbb{I}\!\left[m_{\boldsymbol{\bar{x}}}=1\right]
    \Bigr].
    \end{aligned}
    \label{eq:JSLPPrisk}
\end{equation}

It is proven in~\cite{LehmannV1} that the positive-part versions of these estimators perform even better than the standard versions and dominate them. In the upper part of Fig.~\ref{fig:PPriskplots} the risk for the JSE+, given by \eqref{eq:JSnuPPrisk}, is compared with the risk for the JSE, given by \eqref{eq:JSnurisk} and we see the improved performance near $\nu=0$. The lower part of Fig.~\ref{fig:PPriskplots} compares the risk for JSL and JSL+. Since the Lindley versions of the estimator are not functions of $\boldsymbol{\nu}$, \eqref{eq:JSLrisk} and \eqref{eq:JSLPPrisk} are plotted separately as functions of $\|\boldsymbol{\phi}-\boldsymbol{\bar{\phi}}\|^2$ (a measure of spread for the true parameters) to demonstrate their dominance over the MLE.
The positive part estimators for JSE and JSL, $\boldsymbol{\delta}^\text{JSE+}(\boldsymbol{x},\boldsymbol{\nu})$ and $\boldsymbol{\delta}^\text{JSL+}(\boldsymbol{x})$, both give enhanced estimation. The choice between them depends on whether the centre of the unknown $\phi_i$ values is known or needs to be estimated from the data.

\section{Application to a quantum sensing scheme} \label{sec:quantumSensing}
\begin{figure}[b]
    \centering
    \begin{tikzpicture}
        \tikzset{
            Alice/.style={circle,draw,fill=gray!30,thick,minimum size=1.5cm},
            Bobs/.style={circle,draw,fill=gray!10,thick,minimum size=1.5cm,align=center},
            channels/.style={thick}}
        \node[Alice] (A) at (0,0) {Alice};
        \node[Bobs] (B1) at (5,3){Bob$_1$\\[0.1cm]$\phi_1$};
        \node[Bobs] (B2) at (5,1){Bob$_2$\\[0.1cm]$\phi_2$};
        \node[Bobs] (B3) at (5,-1){Bob$_3$\\[0.1cm]$\phi_3$};
        \node[Bobs] (Bd) at (5,-4){Bob$_d$\\[0.1cm]$\phi_d$};
        \draw[channels] (A) -- (B1);
        \draw[channels] (A) -- (B2);
        \draw[channels] (A) -- (B3);
        \draw[channels] (A) -- (Bd);
        \draw[dashed] (B3) -- node[midway,right]
            {$i=1,2,\ldots,d$} (Bd);
    \end{tikzpicture}
    \caption{A simple networked metrology scheme with a central node, Alice, and $d$ distributed Bobs, where $d\geq3$. Each Bob$_i$ encodes an unknown phase $\phi_i$ locally via a unitary transformation.}
    \label{fig:simplenetworkofsensors}
\end{figure}
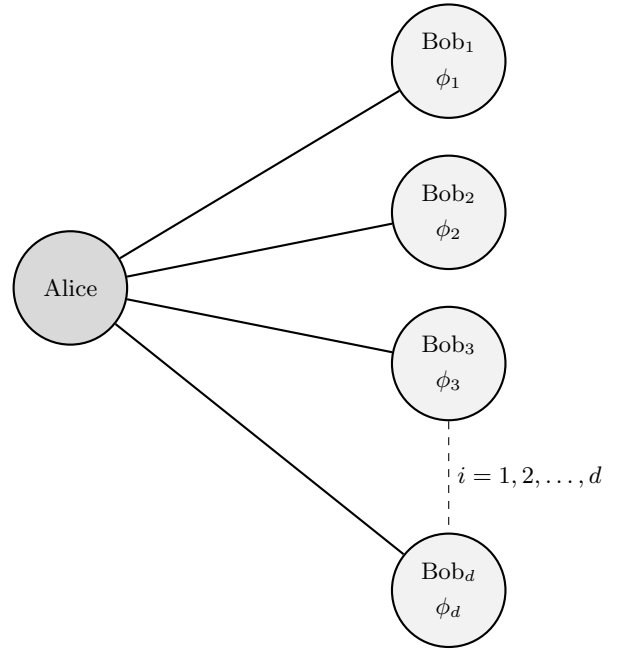

We now consider how the James-Stein estimator and its extensions may be used in networked quantum sensing. Consider a simple network of sensors consisting of a central node (Alice) and several receiving nodes (Bobs), as shown in Fig.~\ref{fig:simplenetworkofsensors}. An independent communication channel runs between Alice and each Bob. Alice's goal is to estimate the phases at each of the Bobs as accurately as possible. This is achieved by sending $n$ known qubits to each Bob that are acted on by a different unitary operator in each channel, which imparts an unknown phase $\phi_i$ onto them. Each Bob then measures each of the $n$-qubits and sends the measurement results through the communication channel to Alice who compiles the data into a single estimate $x_i$ for each $\phi_i$. The measurement scheme can also be made secure to an eavesdropper by Alice sending random qubits (that only she knows) to the Bobs for them to use for their measurements before sending the results back to Alice \cite{Moore2023}. 

We will restrict our attention to a simple noiseless networked quantum metrology scheme where Bob uses only  $\ket{X+}$ eigenstates of the Pauli-X operator, given by 
\[
\ket{X+} = \frac{1}{\sqrt{2}}(\ket{0}+\ket{1}).
\]
After one of the Bobs has encoded phase $\phi$ on it, the state is, 
\[
\ket{X+'} = \frac{1}{\sqrt{2}}(\ket{0}+e^{i\phi}\ket{1}).
\]
The Bob measures in the $X$-basis and the results are $\{+,-\}$. He will observe a $+$ result with probability 
\begin{equation}
    p(+\mid\phi) = \frac{1}{2}(1+\cos{\phi)}
    \label{prob1resultX+}
\end{equation}
and a 
$-$ result with probability 
\begin{equation}
  p(-\mid\phi) = \frac{1}{2}(1-\cos{\phi)}
  \label{prob0resultX+}
\end{equation}
So, in our multi-parameter set-up, each Bob constructs a binomial likelihood function from his measurement results, given by 
\begin{equation}
\mathcal{L}(\phi_i)\propto (1+\cos{\phi_i})^{k_i}(1-\cos{\phi_i})^{n-k_i}
\end{equation}
where $n$ is the number of qubits sent to each Bob and $k_i$ is the observed number of $+$ results for the $i$-th Bob. Once these likelihood functions have been constructed and suitably normalised estimates for the phase values can be extracted, the most common choice of which is the mode of the distribution, corresponding to the maximum likelihood estimator. We will assume that the phases take values in a $[0,\pi]$ range, but all the methods discussed continue to work on a $[0,2\pi)$ interval if standard methods to resolve periodicity (such as using a mixture of different eigenstates of the Pauli matrices) are employed. 

\section{Statistics of the Sensing Scheme}\label{sec:SensingStats}
At the start of our introduction to the JSE we took,  
\[
x_i\mid \phi_i \overset{\text{ind}}{\sim} \mathcal{N}(\phi_i, \sigma^2) \quad \text{where}\ i = 1,2,...,d \quad \text{and}\ d\geq 3.
\]
This contains the following assumptions: 
\begin{enumerate}[itemsep=0pt]
    \item There must be at least three unknown parameters;
    \item The likelihood functions for each parameter must be independent; 
    \item The variances $\sigma^2$ are known and equal (homoskedasticity);
    \item The distributions are normal.
\end{enumerate}
Here we discuss the conditions under which our metrology scheme satisfies these assumptions and how flexible they are in practice. 

The requirement for $d\geq3$ is a hard constraint and the James-Stein estimator is not expected to work if this is not satisfied. This just restricts the class of problems that it can be applied to. The assumption of statistical independence is reasonable because each measurement comes from an independent quantum state. It is well known, e.g. \cite{CASI}, that the James-Stein estimator still performs well when the variances are estimated from the data as opposed to being known exactly. This raises the question as to what variance estimate to use. While sample variances can be used, here we instead choose to use the simple estimate
\begin{equation}
 \hat{\sigma}^2_{CR}=\frac{1}{n}
 \label{CRLB_I=1}
\end{equation}
 since it is known from the Cramer-Rao bound~\cite{Kay} that the variance of an unbiased estimator is bounded by
\begin{equation}
 \sigma^2_{CR}\geq\frac{1}{n\mathcal{I}_n}
 \label{CRLB}
\end{equation}
as $n\to \infty$, where $I_n$ is the Fisher information of each sample. Assuming $I_n = 1$ for all $n$-independent samples, i.e. we assume that the Bobs are making perfect measurements, this reduces to \eqref{CRLB_I=1}. The Fisher information for a single measurement is usually known in an experiment, so if it differs from unity we would simply use expression \eqref{CRLB} with the appropriate denominator. The limit \eqref{CRLB} is independent of the value of the unknown phase, so the variances of all the likelihood functions will be well estimated by this value if there is a reasonable number of qubits available and each Bob uses the same number. Note again that variance estimates of this form, which will be used throughout this paper for simplicity, are asymptotic results that hold in large data. This is justified when operating in a Gaussian regime (which will be discussed further below), and continues to work outside of this, but the analyst could always choose to use the full form of the classical Fisher Information, that is \eqref{CRLB} with 
\begin{equation}
    \mathcal{I} = \int d\boldsymbol{x} (\partial_\phi\mathcal{L}(\phi,\boldsymbol{x}))^2/\mathcal{L}(\phi,\boldsymbol{x})
    \label{ClassicalFI}
\end{equation}
instead. 

We now discuss the assumption of normality. The fact that the likelihood functions in our scheme are binomial (as opposed to normal) has a few consequences, all of which are fairly easily dealt with. Firstly, the binomial distribution can be well approximated by a normal when the condition 
\begin{equation}
    np, n(1-p) > 5
    \label{NormalApproxToBinoCondition}
\end{equation}
is satisfied \cite{ClarkeCooke}. Whilst this condition will not always be satisfied in our scheme, there are entirely geometric arguments for Stein-type estimators presented in~\cite{FSW,BrownAndZhao} which are made without any assumption of normality. This suggests that in practice we should continue to expect the JSE to work well even in the binomial regime, despite the fact that we should only expect the analytical MSE functions presented earlier to hold when this condition is satisfied. The JSE has been applied to the estimation of several binomial probabilities before, most famously by Efron and Morris in \cite{Efron75} and \cite{CASI}. 

\begin{figure}[b]
    \centering
    \includegraphics[width=\columnwidth]{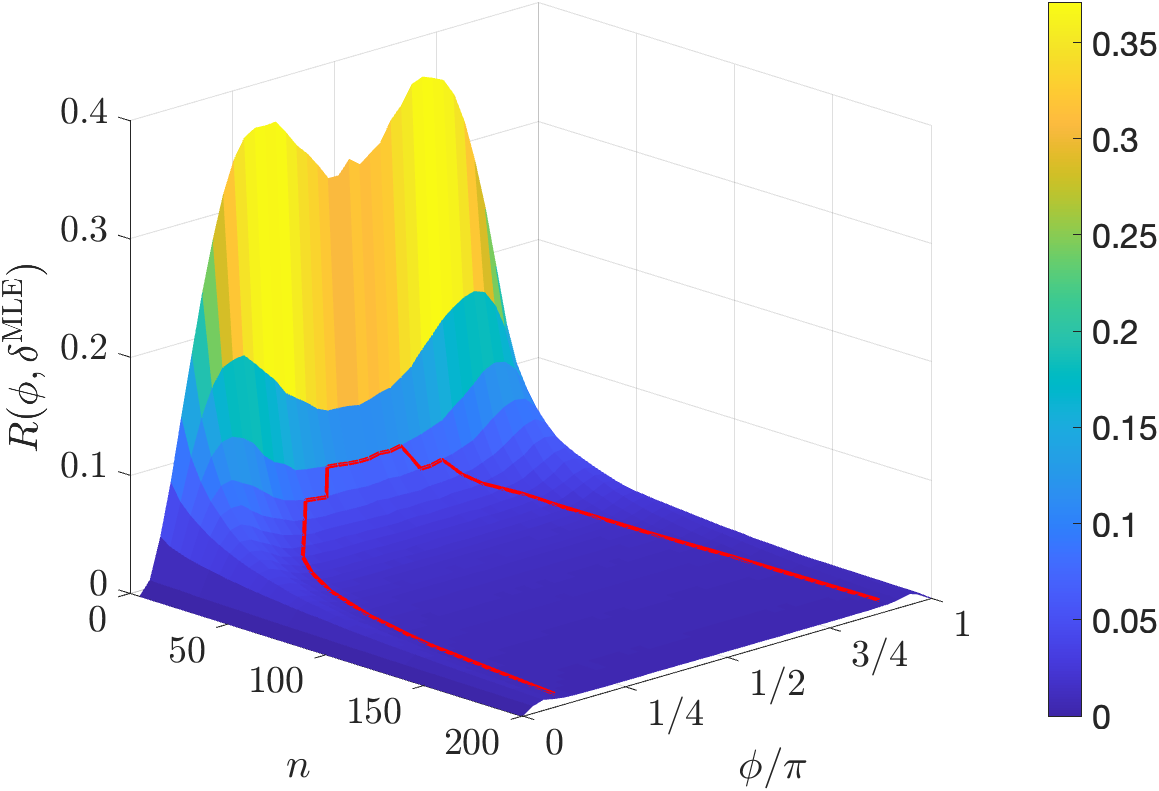}
    \caption{MSE for the MLE of a single phase ($d=1$) as a function of the true value, $\phi$, and the number of measurements, $n$. The red line encloses the region of approximate normality.}
    \label{fig:Figure5}
\end{figure}

A further consequence of the likelihood functions in our scheme being binomial is the boundedness of the support. By observing the form of the variance in the estimation of a binomial probability $p_i$ \cite{Carlin},
\begin{equation}
    \sigma^2_i = \frac{1}{n}{p_i}(1-{p_i})
    \label{BinomialVariance}
\end{equation}
it is easy to see that if the true probability in our measurement scheme takes the value of the lower or upper limit exactly, that is $p_i = 0$ or $p_i = 1$ (\eqref{prob1resultX+} gives the corresponding phase values as $\phi_i=0$ and $\phi_i=\pi$), the binomial measurements that are observed become completely deterministic and the error in the MLE vanishes, i.e.  the MLE predicts the true value exactly~\cite{LehmannV2}. This behaviour manifests itself at the edges of Fig.~\ref{fig:Figure5}, where the MSE of the maximum likelihood estimate for a single phase is plotted against $n$ and $\phi\in[0,\pi]$. The region where approximate normality holds (determined by \eqref{NormalApproxToBinoCondition}) and so the region where the MSE is approximately independent of the true value and 
\begin{equation}
    R(\phi,\delta^{MLE})\approx \hat{\sigma}^2_{CR}
\end{equation}
is enclosed by the red line. In the cases where $p_i = 0$ or $p_i = 1$ the JSE cannot reduce the error of the MLE any further; applying any shrinkage would be sure to make the estimate worse, so the dominance of the JSE cannot hold over the entire support~\cite{Johnson71}. However, we can be confident that the James-Stein will continue to dominate the MLE over most of the support~\cite{Gutmann82a}. In practice, this simply means that if, when gathering data, it looks like any of the true phases lie close to values that gives probability $0$ or $1$, the MLEs for those parameters should be excluded from the shrinkage procedure and the JSE only applied to the remaining phase estimates.

The dependence of the MSE on the true value (as seen in Fig.~\ref{fig:Figure5}) leads to another side-effect that differs from a normal distribution. The mean and variance of a normal distribution are independent \cite{ClarkeCooke} and so, if the spread of the true values is fixed (e.g. $\tau^2$ in \eqref{TVdistributionNormal}), the exact values the unknown means take do not affect the resulting MSE; the risk function for the MLE is always given by \eqref{eq:MLErisk}. Given the nature of the likelihood functions in our scheme, however, we should not expect this to be the case. As can be seen from Fig.~\ref{fig:Figure5}, different values of $\phi$ contribute different amounts to the total MSE, so the resulting MSE curves will depend on both the spread of the true phases and the specific values they take. As a result, for any given fixed spread of true phases we expect there to be a set of possible risk functions we could obtain, dependent on the true phases themselves. This is unlike the standard Gaussian setting where there is only one possible MSE curve for a particular spread of true values, as seen in Fig.~\ref{fig:Figure2} and Fig.~\ref{fig:PPriskplots}. 
\vspace*{1cm}

\section{Simulations for the estimation of several phases}\label{sec:Simulations}
The range of different MSE curves can be obtained from simulations. In this section, we show the results of Monte-Carlo simulations of the MSE for the MLE and JSE.
Importantly, these simulations show that, despite the probability distributions not being normal, all the expected behaviour of the estimators in different limits continue to hold, and that performance comes out as predicted by the analytical Gaussian results in the expected regimes. The results show that the general performance of the Stein-type estimators are remarkably robust to deviations from the model assumptions, and that the range of estimation problems to which they can be applied may be wider than expected.

\begin{figure*}[t]
\centering
\begin{subfigure}{\columnwidth}
    \centering
    \caption*{\hspace*{-20pt}(a) $r=12,\ \tau^2=0$}
    \hspace*{-40pt}
    \includegraphics[width=\columnwidth]{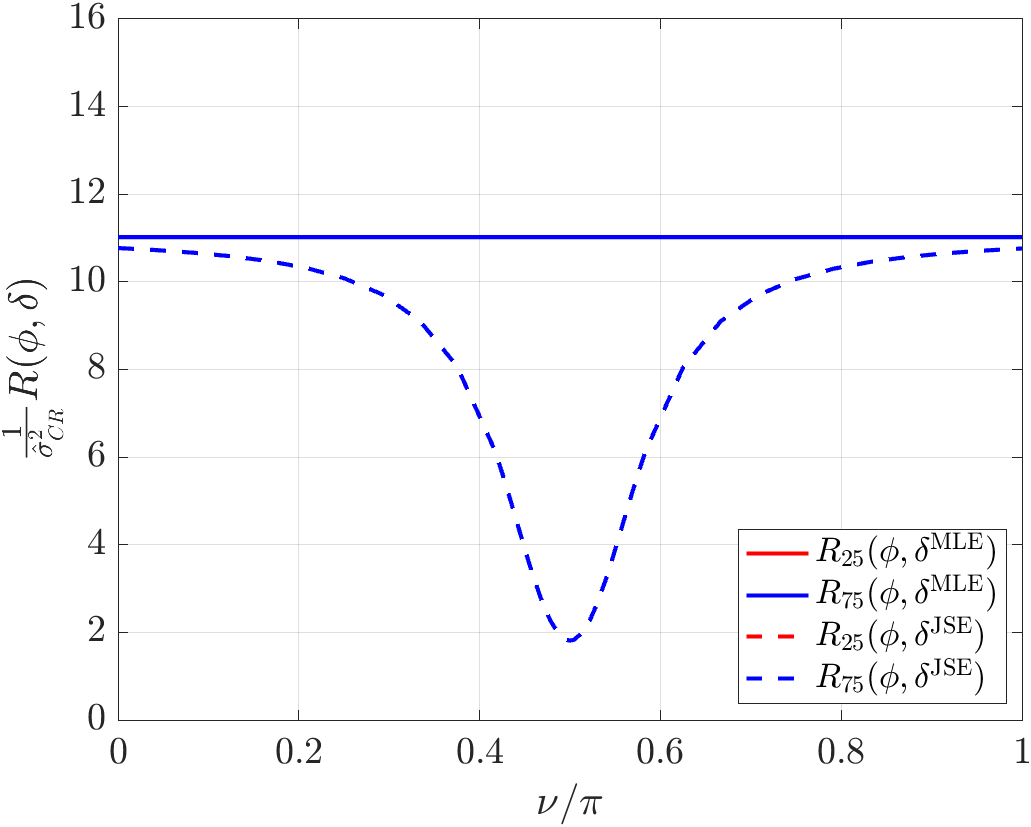}
\end{subfigure}
\hfill
\begin{subfigure}{\columnwidth}
    \centering
    \caption*{\hspace*{-20pt}(d) $r=60,\ \tau^2=0$}]
    \hspace*{-40pt}
    \includegraphics[width=\columnwidth]{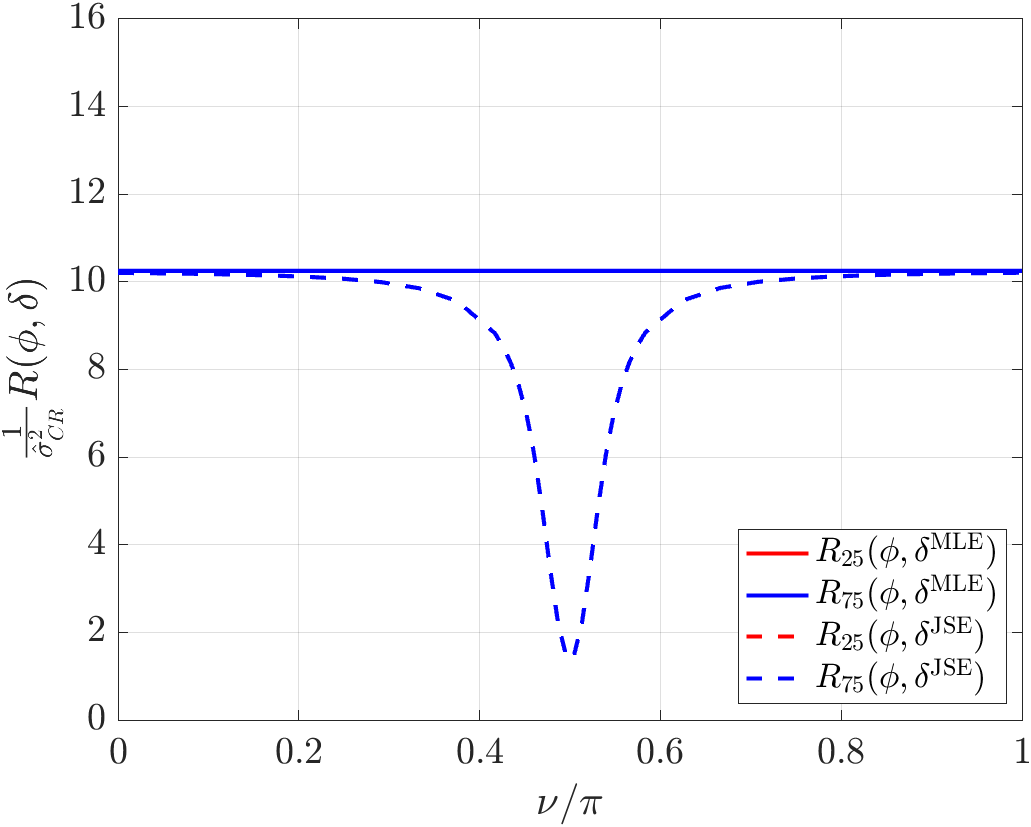}
\end{subfigure}
\\[10pt]
\begin{subfigure}{\columnwidth}
    \centering
    \caption*{\hspace*{-20pt}(b) $r=12,\ \tau^2=\frac{1}{24}\pi^2$}
    \hspace*{-40pt}
    \includegraphics[width=\columnwidth]{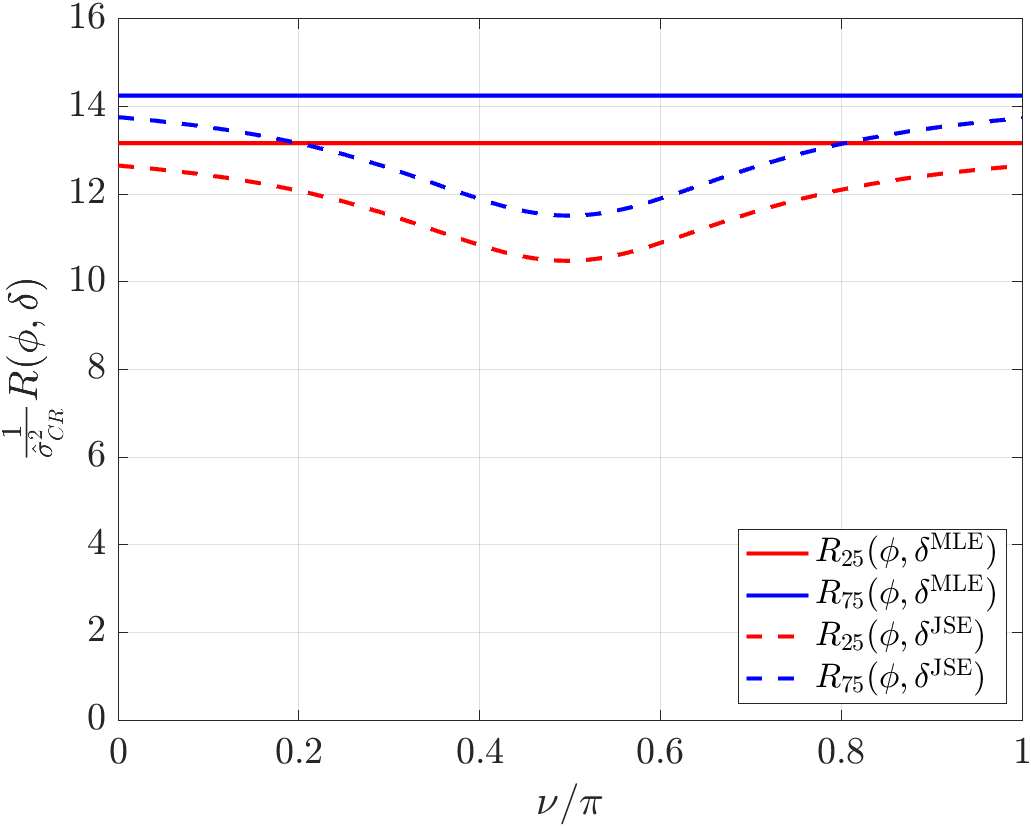}
\end{subfigure}
\hfill
\begin{subfigure}{\columnwidth}
    \centering
    \caption*{\hspace*{-20pt}(e) $r=60,\ \tau^2=\frac{1}{24}\pi^2$}
    \hspace*{-40pt}
    \includegraphics[width=\columnwidth]{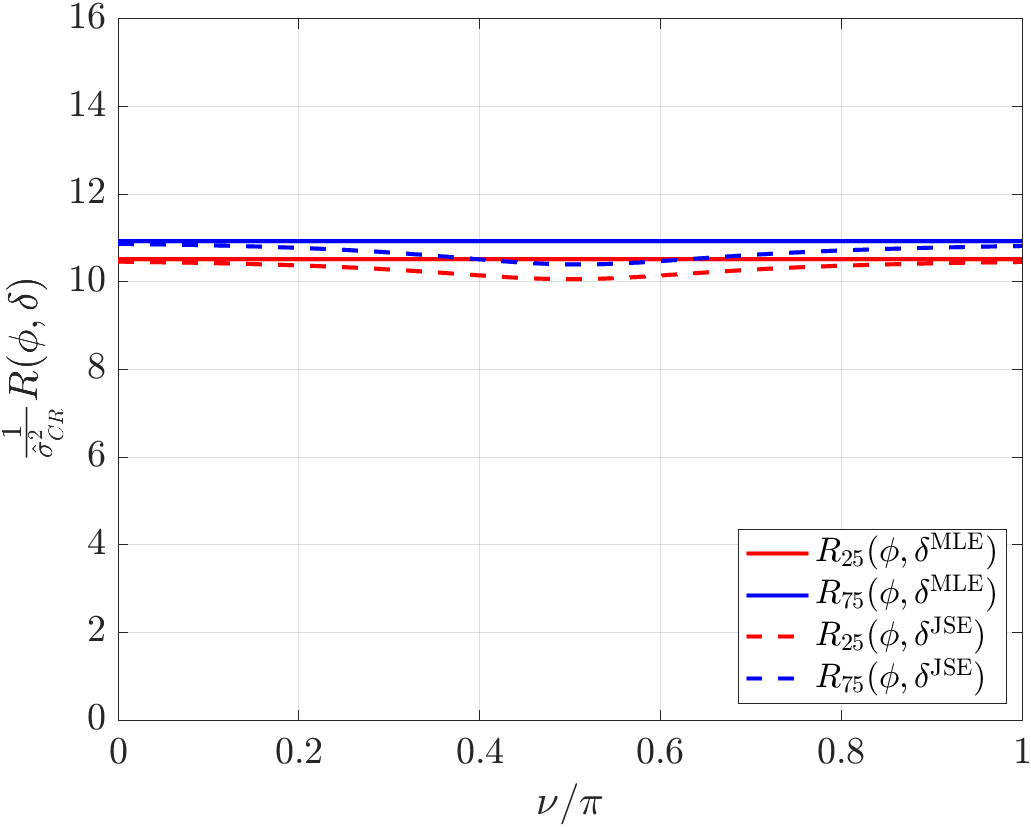}
\end{subfigure}
\\[10pt]
\begin{subfigure}{\columnwidth}
    \centering
    \caption*{\hspace*{-20pt}(c) $r=12,\ \tau^2=\frac{1}{12}\pi^2$}
    \hspace*{-40pt}
    \includegraphics[width=\columnwidth]{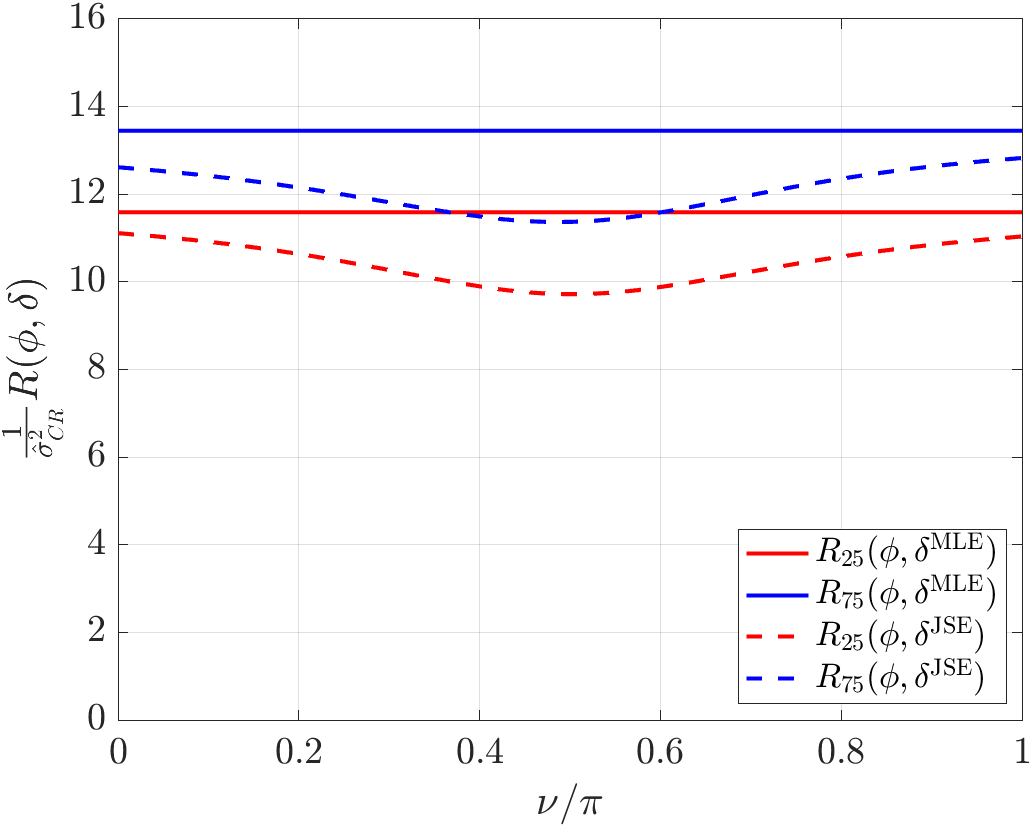}
\end{subfigure}
\hfill
\begin{subfigure}{\columnwidth}
    \centering
    \caption*{\hspace*{-20pt}(f) $r=60,\ \tau=\frac{1}{12}\pi^2$}
    \hspace*{-40pt}
    \includegraphics[width=\columnwidth]{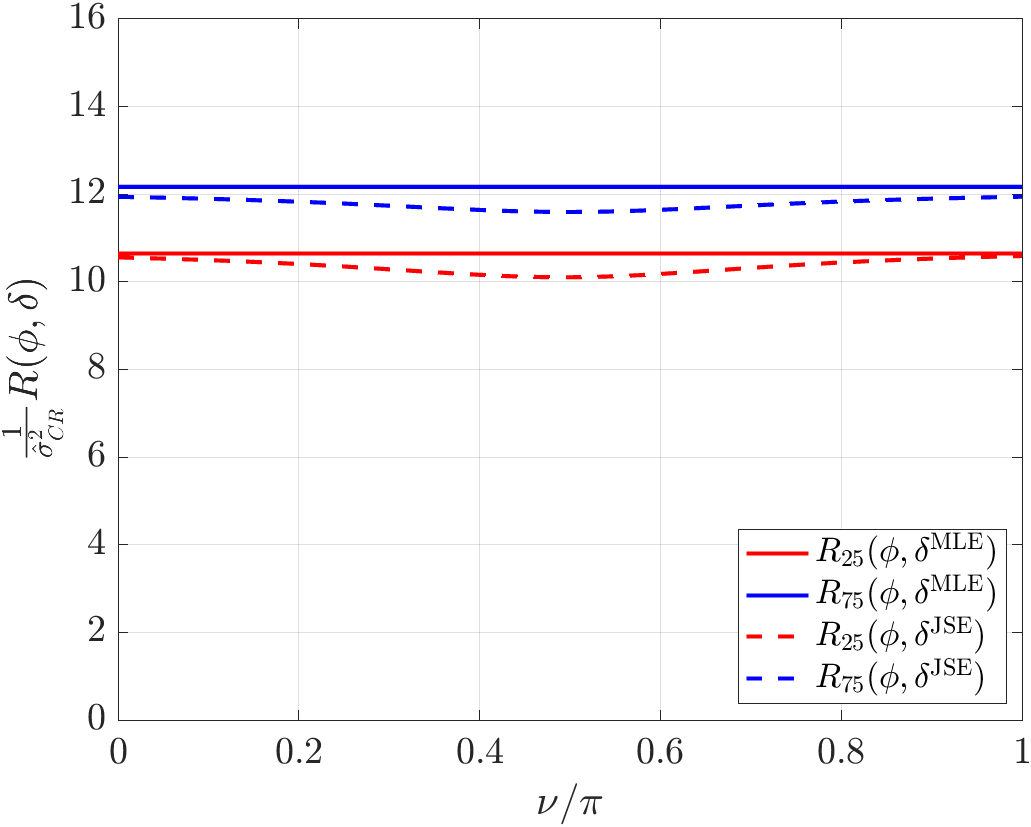}
\end{subfigure}
\caption{The 25th and 75th percentile MSE curves obtained for the estimation of $d=10$ phases using $\ket{X+}$ states for two choices of $r$ and three choices of $\tau^2$. We see that the JSE never performs worse than the MLE and performs much better when the total resources, $r$, and the variance of the true values, $\tau^2$, are low. The red and blue curves lie on top of one another in (a) and (d).}
\label{fig:X+bucketplots}
\end{figure*}

The unknown phases for the simulations are drawn from a continuous uniform distribution between a lower bound $a$ and upper bound $b$, i.e. 
\begin{equation}
    \phi_i \sim \mathcal{U}[\frac{\pi}{2}-c,\frac{\pi}{2}+c].
    \label{TVdistributionContUniform}
\end{equation}
We denote the variance of this distribution as $\tau^2$, which is in general unknown and is bounded from below by $0$ when $c = 0$ and from above by $\frac{1}{12}\pi^2$ when $c = \frac{\pi}{2}$. We denote this maximum variance $\tau^2_{max}$. For a fixed value of $\tau^2$, we generate many combinations of phases that satisfy the mean and variance constraints. We make MLEs of the $d$-phases and apply the positive-part James-Stein estimator \eqref{eq:JSnuPP} to them. The observed MSE for the two estimators for each set of true values is recorded and plotted as a function of $\boldsymbol{\nu}$, again where each element of $\boldsymbol{\nu}$ is chosen to be identical. Thus, we have a range of MSE curves that we could obtain for a set of true values of fixed mean $\pi/2$ and variance $\tau^2$. For simplicity, we choose to plot the 25th and 75th percentile MSE curves (averaged over $\nu$) of those obtained via simulation. We denote these $R_{25}$ and $R_{75}$ respectively.  Throughout the remainder of this paper we will plot the scale-invariant risk \cite{FSW}, where we have divided the MSE by the variance estimate so that the resulting risk plots are given in units of variance. For consistency with the later discussion using quantum enhanced states,  we
define the resources $r$ sent to each individual Bob as the number of qubits $n$ multiplied by the number of particles within each qubit $N$, so that $r = n\cdot N$. For $\ket{X+}$ we have $N=1$, so $r=n$. 
Fig.~\ref{fig:X+bucketplots} shows the scale-invariant risk as a function of $\nu$ for $\tau^2$=$0$, $\tau^2$=$\pi^2/24$ and $\pi^2/12$ and for $r=12$ and $60$. We see that, as predicted by the analytical expressions the JSE  performs better when the true values are closer to the chosen centre point, that is when $\tau^2$ is low. This can be seen by comparing plots (a)-(c) or (d)-(f), which show the results for increasing $\tau^2$ as resources $r$ remain fixed. It can also be seen that the effectiveness of the JSE diminishes as $r$ increases, as predicted by \eqref{JSrisktendstoMLErisk} - this is easily visible by comparing plots (a) and (d), (b) and (e), or (c) and (f), which show the MSE curves for increasing $r$ and fixed variance.  
\clearpage
These results cast the Stein estimators as being particularly useful in low data or cryptographic scenarios, where it is of primary interest to share as few bits of information as possible between parties. In practice the analyst would only shrink the MLEs towards a single point as opposed to a full range of $\nu$ choices. Since the value of $\tau^2$ is unknown, these plots serve to demonstrate that it remains possible to shrink towards \textit{any} choice of $\nu$ and the Stein estimator still performs to at least to the same level as the maximum likelihood estimator, as predicted by \eqref{eq:JSnuPPrisk}, and sometimes does much better, most notably when the choice of $\nu$ is indeed the centre point of the true phases. Again if this centre point is unknown it can be estimated as the sample mean of the group of MLEs with little loss in performance (recall the Lindley estimators presented earlier, \eqref{eq:JSL} and \eqref{eq:JSLPP}).

We now consider the realistic scenario where you shrink towards a single centre point in more detail, in which we take the centre point to be $\nu=\pi/2$ for simplicity. We vary $\tau^2$ between $0$ and its maximum value, simulating many sets of true values that satisfy the mean and variance constraints. MLEs of these phases are created and the positive-part James-Stein estimator \eqref{eq:JSnuPP} is applied to them. The MSE of the estimates is recorded. We then find the differences between the MSE of the MLEs and the JSEs and calculate the average difference for each value of $\tau^2$ - we denote this average risk difference by $\bar{R}_{JSE-MLE}$. 

\begin{figure}[h]
    \centering
    \begin{subfigure}{0.45\textwidth}
    \hspace{1cm}
    \hspace*{-40pt}
    \begin{minipage}{\linewidth}
        \centering
        \caption*{\hspace*{+20pt}(a) $r=12$}
        \includegraphics[width=\linewidth]{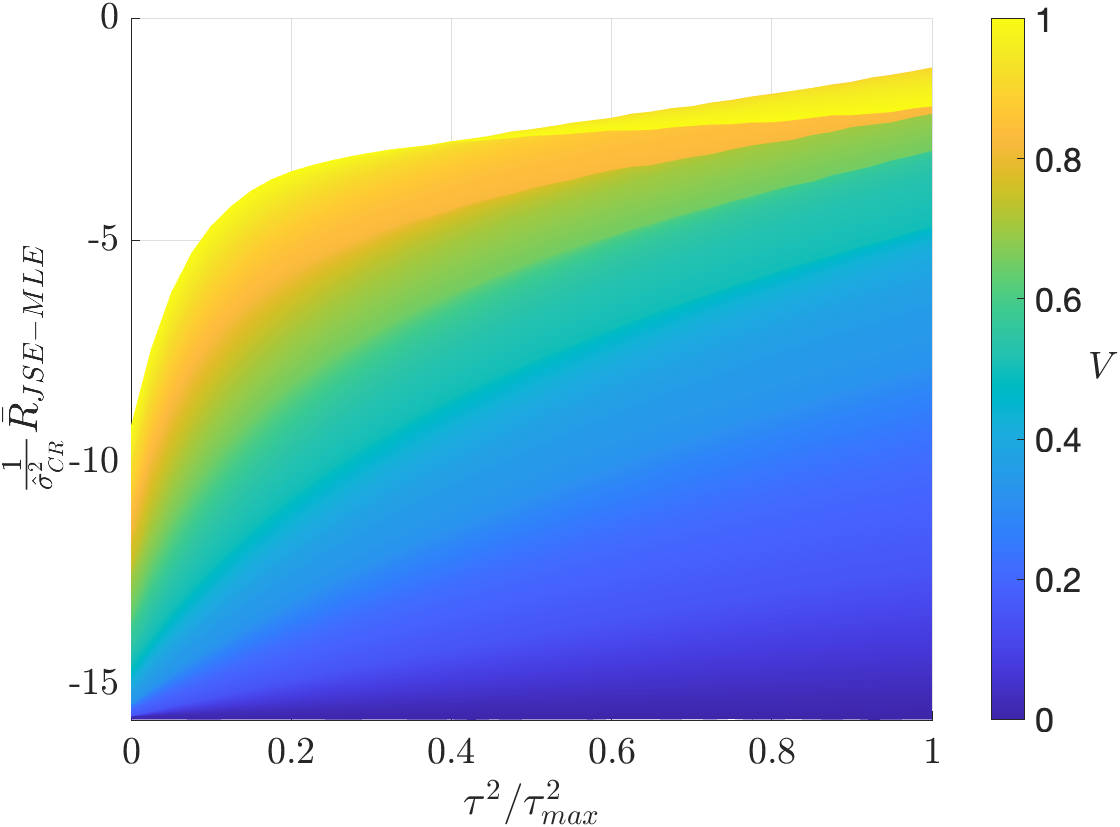}
    \end{minipage}
    \end{subfigure}
    \\[10pt]
    \hfill
    \begin{subfigure}{0.45\textwidth}
        \hspace{0.5cm}
        \hspace*{-40pt}
        \begin{minipage}{\linewidth}
            \centering
            \caption*{\hspace*{+20pt}(b) $r=60$}
            \includegraphics[width=\linewidth]{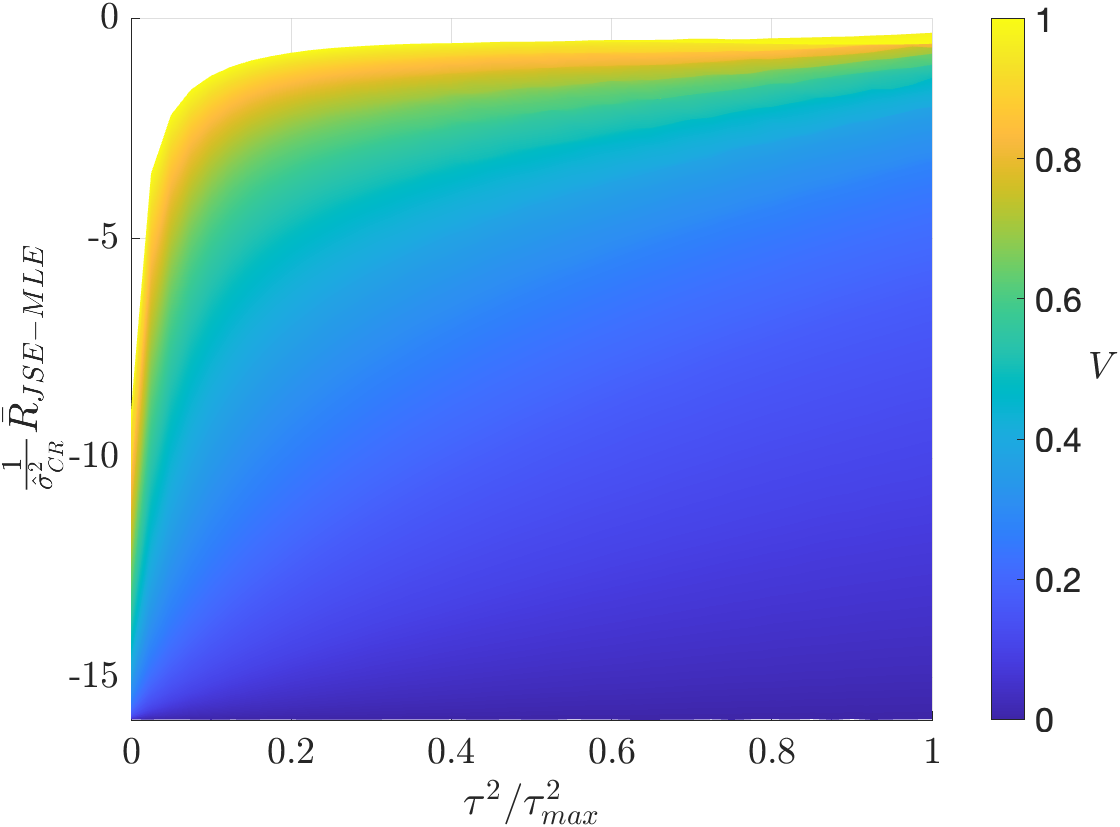}
        \end{minipage}
    \end{subfigure}
    \caption{Average difference in risk at $\nu = \pi/2$ between the MLE and JSE
    for (a) $r=12$ and (b) $r=60$. The JSE outperforms the MLE by a significant
    amount even when the true phases are maximally spread, i.e.\ $\tau^2/\tau_{max}^2 = 1$. This is especially the case as $V\to0$.}
    \label{fig:X+RDplots}
\end{figure}

So far we have only considered the noiseless setting of single qubit metrology. In practice however, there is always some level of noise present due to imperfections in hardware and contamination from the environment. This raises the question 
as to how well the shrinkage performs in the presence of some noise. A simple noise model comes from adding a so-called \enquote*{visibility parameter} $V$ onto our earlier expressions, which from \cite{Roccia2018,Zych2011} become 
\begin{equation}
    p(\pm|V,\phi) = \frac{1}{2}(1\pm V \cos(\phi)),\quad\text{where } V \in [0,1].
    \label{visibilityProbabilities}
\end{equation}
The noiseless scenario considered earlier thus represents the case when $V=1$, with noise increasing as $V\to0$. Fig.~\ref{fig:X+RDplots} plots the scaled average risk difference $\frac{1}{{\hat{\sigma}}^2_{CR}}\bar{R}_{JSE-MLE}$ against $V$ and $\tau^2/\tau^2_{max}$ for (a) $r=12$ and (b) $r=60$. Note that the shrinkage estimator 
works particularly well not only when both $r$ and $\tau^2$ are low (as might be expected), but also when $V$ is closer to zero. We see significant reductions in average risk when $V$ is sufficiently low, irrespective of the value of $\tau^2$. This constitutes a scenario where added noise actually makes the shrinkage estimation \textit{more effective}, which is understandable by considering the form of \eqref{visibilityProbabilities}. As $V\to0$, $p(\pm|V,\phi)\to0.5$, which is the region of probability space where the normal approximation to the binomial holds most strongly. In addition to this, from \eqref{JSrisktendstoMLErisk}, we should expect the Stein estimators to perform particularly well compared to the MLE when the uncertainty in the MLE is higher. As $V\to0$, the uncertainty in the MLE increases. The more noise there is in the estimation problem, the less we should trust the MLE and the greater the positive impact we should expect applying a Stein-type estimator to have on reducing the MSE. This builds on the ideas introduced in \ref{sec:SensingStats} where it was shown that we should expect the shrinkage estimator to perform differently depending on the amount of uncertainty in the initial MLE of the phases. Note that in the context of the visibility parameter Fig.~\ref{fig:Figure5} was simulated in the noiseless case, i.e. with $V=1$, so there would be a slightly different version of this plot for each value of $V$. 
\vspace*{-5pt}
\clearpage
\section{Entanglement-enhanced quantum measurements} \label{sec:quantumEnhanced}
To date, JSE and its extensions have barely appeared in the field of quantum metrology. We have shown above that they can be a valuable addition to networked schemes with separable qubits by reducing the MSE through data analysis without additional experimental requirements. An interesting question is whether the JSE can also be applied to schemes utilizing entanglement so that we can effectively win twice through both the entanglement-enhanced precision and the JSE improvement in the MSE. Here we present an example of a scheme that shows this to be the case.  

\begin{figure}[h]
    \centering
    \begin{subfigure}{0.45\textwidth}
    \hspace{1cm}
    \hspace*{-40pt}
    \begin{minipage}{\linewidth}
        \centering
        \caption*{\hspace*{+20pt}(a) $r=12$}
        \includegraphics[width=\linewidth]{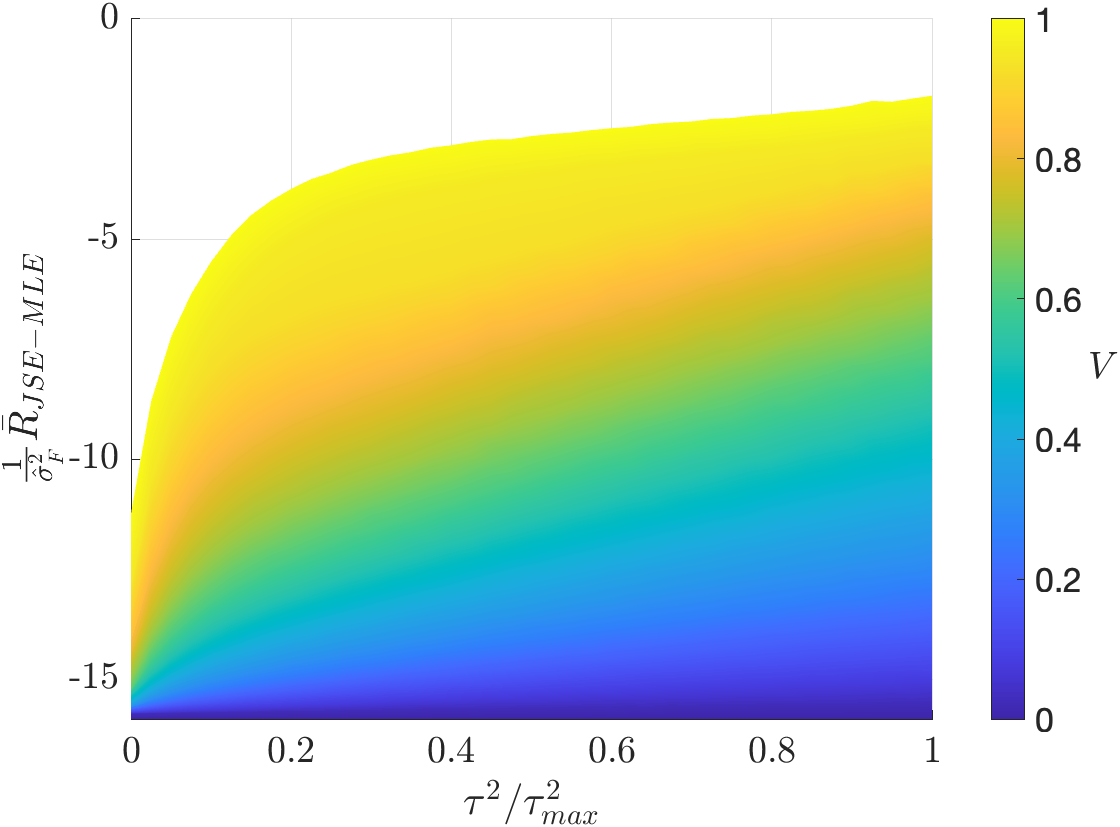}
    \end{minipage}
    \end{subfigure}
    \\[10pt]
    \begin{subfigure}{0.45\textwidth}
    \hspace{1cm}
    \hspace*{-40pt}
    \begin{minipage}{\linewidth}
        \centering
        \caption*{\hspace*{+20pt}(a) $r=60$}
        \includegraphics[width=\linewidth]{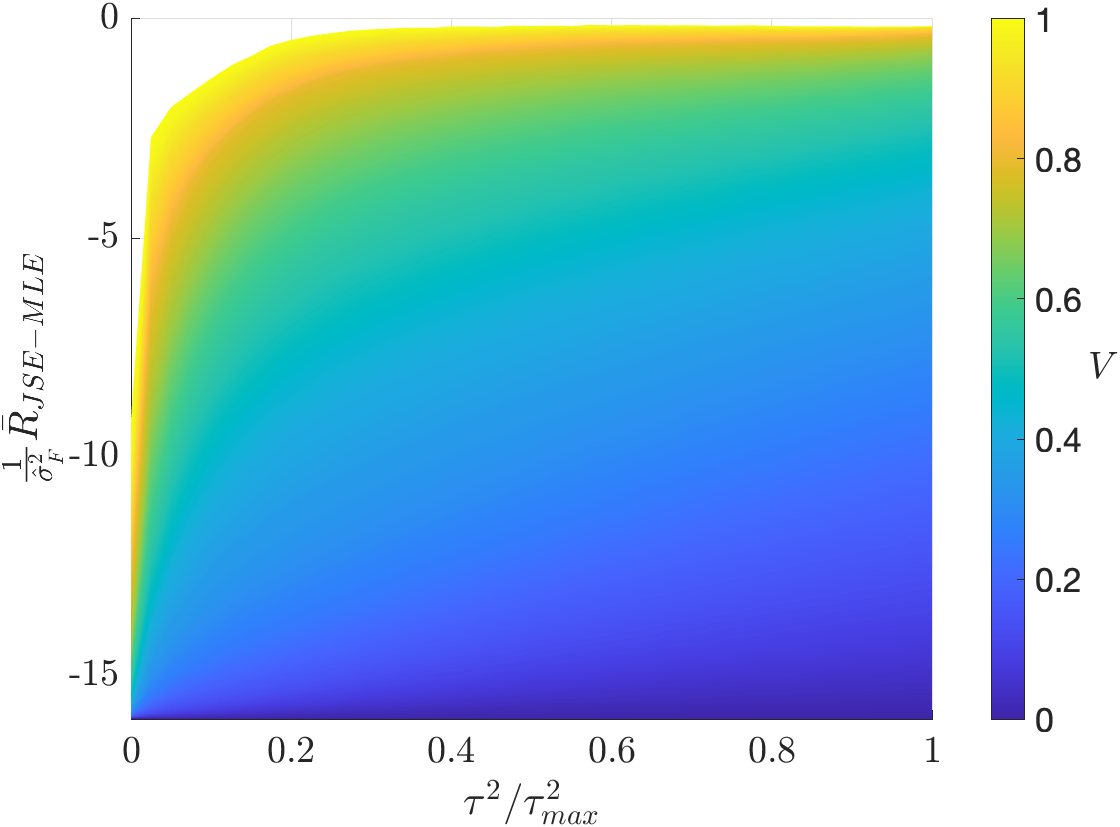}
    \end{minipage}
    \end{subfigure}
    \caption{Average risk differences with an equal mixture of the $N = 1,2,3$ NOON states for (a) $r=12$ and (b) $r=60$. Similar behaviour is displayed to that of the entanglement-free scheme shown in Fig.~\ref{fig:X+RDplots}.}
    \label{fig:NOONStatesRDplots}
\end{figure}

Consider the $N$-particle NOON states~\cite{Kok2002} of size $N = 1, 2, 3$, while keeping the rest of the scheme the same as presented in Sec.~\ref{sec:quantumSensing}. After encoding the phase, $\phi$, these states have the form
\[
\ket{{\rm NOON}} = \frac{1}{\sqrt{2}}(\ket{N,0}+e^{iN\phi}\ket{0,N}).
\]
The probabilities of observing the results of $\{+,-\}$ become
\begin{equation}
    p(\pm\mid\phi) = \frac{1}{2}(1\pm\cos{N\phi)}
    \label{probresultNOON}
\end{equation}
Each Bob can use their measurement counts to construct a likelihood function as in Sec.~\ref{sec:quantumSensing}. This takes the form
\begin{equation}
    \mathcal{L}_{N}(\phi_i) \propto (1+\cos{N\phi_i})^{k_i}(1-\cos{N\phi_i})^{n-k_i},
    \label{LikelihoodFunctionForNOONs}
\end{equation}
where $n$ is the number of qubits available for each Bob and $k_i$ is the observed number of $+$ results for the $i$-th Bob. If we assume that each Bob is making perfect measurements, the bound for the variance of this likelihood function for a single NOON state of size $N$ has Heisenberg scaling \cite{Giovannetti2004}, 
\begin{equation}
    \hat{\sigma}^2 \geq \frac{1}{nN^2}.
    \label{HeisenbergScaling}
\end{equation}
and the resource count is $r=n \cdot N$. The likelihood (\ref{LikelihoodFunctionForNOONs}) is $N$-fold periodic, so can give rise to an identifiability issue as to which value of the parameter is correct. As such we consider the case where each Bob uses a mixture of different sized NOONs, $N=1,2,3$, to form three independent likelihoods for the same phase. These are then multiplied together to form a final single-peaked likelihood,
\begin{equation}
    \mathcal{L}(\phi_i) \propto 
    \mathcal{L}_{N=1}(\phi_i)\cdot
    \mathcal{L}_{N=2}(\phi_i)\cdot
    \mathcal{L}_{N=3}(\phi_i).
\label{FinalLikelihoodFunctionForNOONs}
\end{equation}
If we assume that each of the constituent likelihoods $\mathcal{L}_{N=1}$, $\mathcal{L}_{N=2}$ and $\mathcal{L}_{N=3}$ contains peaks that are Gaussian and substitute in variances of the form \eqref{HeisenbergScaling}, the final variance, $\sigma_F^2$, of $L(\phi_i)$ is bounded from below by,
\begin{equation}
   \hat{\sigma}^2_{F} \geq \frac{1}{n\sum_{j=1}^{J}{N_{j}^2}}
   \label{VarFinal}
\end{equation}
where, in this case, $J=3$ and the resource count is $r=6n$. We then apply the James-Stein positive-part estimator \eqref{eq:JSnuPP} (now using \eqref{VarFinal} as the variance  estimate) to the MLEs of the likelihood functions for the $d=10$ phases. As before, we take these phases to be generated from the continuous uniform distribution \eqref{TVdistributionContUniform}. In exactly the same style as Sec VI, the results of these simulations are shown in Fig.~\ref{fig:NOONStatesRDplots}. 
For comparison with these earlier plots, the same resource counts (a) $r=12$ and (b) $r=60$ have been used. We observe similar results, in particular that \eqref{eq:JSnuPP} significantly reduces MSE when $r$, $\tau^2$ and $V$ are low. We would like to point out that $\hat{\sigma}^2_F$ is smaller than $\hat{\sigma}^2_{CR}$ for the same resource count, owing to the enhancement in precision achieved using entangled states. This means that whilst the \textit{proportional} benefit of using the James-Stein is similar in both the entangled and unentangled cases, the \textit{absolute} reduction in MSE would be lower in the entangled case since we are starting from a point of higher precision. This also means that we gain an advantage twice -- once from the entanglement and once from the shrinkage, which raises the question as to which effect gives a greater improvement in precision. This is something that could be explored in future work. 

\section{Conclusions}
In this work, we have shown that James–Stein shrinkage estimation and its extensions can provide significant reductions in mean-squared error in distributed quantum sensing without requiring any additional quantum resources or modifications to the underlying experimental architecture. By applying shrinkage methods at the data-processing stage, we demonstrated that substantial performance gains can be achieved in the estimation of multiple phases, particularly in the practically important regimes of limited data and when noise is present.

We analysed the behaviour of the positive-part James–Stein estimator within a networked quantum sensing framework, identifying the conditions under which these methods are most effective. Through Monte Carlo simulations, we showed that the advantages of shrinkage estimation remain remarkably robust even when the underlying assumptions of the ideal Gaussian model are relaxed and the measurement statistics are intrinsically binomial. In particular, the estimator continues to outperform standard maximum-likelihood estimation across a broad region of parameter space, especially when the available measurement resources are low and there is noise present in the estimation problem. Similar results are obtained if the Lindley version of the estimator is used instead.

We also demonstrated that shrinkage estimation can be combined naturally with quantum-enhanced sensing protocols based on entangled NOON states. In this setting, the improvement from the JSE acts in addition to the enhancement already provided by entanglement, yielding a dual advantage in precision. This suggests that classical statistical techniques and quantum resources are complementary tools that can be jointly optimised within a unified metrological framework. The application of shrinkage estimators to entangled schemes will be explored further in future work. The authors also suggest that applying shrinkage in the unequal variance case, or to the case where different numbers of qubits $n$ are sent to each Bob, could prove to be fruitful research directions. 

More broadly, this work highlights the importance of the often-overlooked data-processing stage in quantum metrology. While considerable effort has been devoted to the optimisation of quantum states, measurements, and noise mitigation strategies, comparatively little attention has been paid to the role of advanced statistical inference techniques. Our results indicate that there exists a rich set of underexplored methods from classical statistics that may offer substantial practical advantages for quantum sensing and metrology. We anticipate that closer interaction between the fields of quantum metrology, statistical inference, and machine learning could open new directions for improving quantum-enhanced measurements and distributed sensing networks.
\section*{Acknowledgements}
LAR and JAD gratefully acknowledge funding for this project from the United Kingdom's DSTL. SWM acknowledges support from the ANR project EQUINE (ANR-23-QUAC-0001). The authors thank Nathan Shettell for introducing them to the idea of the James-Stein estimator.

\end{document}